\documentstyle[psfig,twocolumn,prl,aps]{revtex}
\begin{document}\draft
\twocolumn[\hsize\textwidth\columnwidth\hsize\csname @twocolumnfalse\endcsname
\title{Geometrical model for a particle on a rough inclined surface}
\author{ Giovani L. Vasconcelos and J. J. P. Veerman}
\address{
Departamento de F\'{\i}sica, Universidade Federal de Pernambuco,
50670-901, Recife, Brazil.}
\date{To appear in Physical Review E, 01 May 1999}
\maketitle 
\begin{abstract}
A simple geometrical model is presented for the gravity-driven motion
of a single particle on a rough inclined surface. Adopting a simple
restitution law for the collisions between the particle and the
surface, we arrive at a model in which the dynamics is described by a
one-dimensional map. This map is studied in detail and it is shown to
exhibit several dynamical regimes (steady state, chaotic behavior, and
accelerated motion) as the model parameters vary.  A phase diagram
showing the corresponding domain of existence for these regimes is
presented. The model is also found to be in good qualitative agreement
with recent experiments on a ball moving on a rough inclined line.
\end{abstract}
  \vskip1pc] 

\section{Introduction}

Several experimental studies \cite{jan,riguidel1,riguidel2,ristow}
have recently been conducted on the problem of a single ball falling
under gravity on a surface of controlled roughness.  These works have
revealed interesting new aspects of granular dynamics that are not yet
fully understood.  Three distinct dynamical regimes have been
identified \cite{jan,riguidel1,riguidel2,ristow} as the tilting angle
increases. For small inclinations there is (i) a decelerated regime
where the ball always stops, then comes (ii) an intermediate regime
where the ball reaches a steady state with constant mean velocity, and
for larger inclinations the ball enters (iii) a jumping regime.
Computer simulations
\cite{riguidel1,riguidel2,ristow,batrouni1,batrouni2} have confirmed
these results, particularly those concerning regimes (i) and (ii).  A
theoretical model \cite{ancey} has also been proposed in which
steady-state solutions (but no detailed dynamics) can be obtained
analytically. More recently, a one-dimensional map \cite{valance} has
been introduced to study the jumping regime. This map in its simplest
version is linear, and to obtain non-linear behavior one has to vary
spatially the properties of the rough surface \cite{valance}, in which
case the model become inaccessible analytically.

In this Paper we present a model for a single particle moving under
the action of gravity on a rough surface of specified shape.  Within
this setting we will give a detailed analytical description of all
possible dynamical regimes.  Although the model we study is
simplified, its predictions are in good qualitative agreement with the
experimental findings.

Roughly speaking, our conclusions are as follows. There is (i) a sharp
transition (as the surface inclination increases) from a regime of
bounded velocity to one of accelerated motion. Within the region of
bounded velocity various dynamical regimes are possible.  First there
is (ii) a range of inclinations for which the dynamics always has a
unique attractor. For higher inclinations two other phases exist:
(iii) a region where we have co-existing attractors for the dynamics
and (iv) a region where instabilities give rise to chaotic behavior.
For a fixed (sufficiently large) inclination a transition to the
chaotic region will take place as the nature of the collisions between
the particle and the surface becomes highly inelastic.  Although our
results are derived here in the context of a simple collision rule, it
can be shown \cite{companion} that they remain valid for a wide class
of restitution laws.

The paper is organized as follows. In Sec.\ II we describe the model
and study in detail its dynamical properties. The main results of this
Section are then summarized in the phase diagram shown in Fig.\
\ref{fig:phase}. In Sec.\ III we carry out a comparison between the
model predictions and the experimental findings. In particular, we
argue that the jumping regime seen in the experiments might correspond
to a true chaotic motion, as predicted by the model. Finally, in Sec.\
IV we collect our main conclusions and present further discussions.

\section{the model}

\begin{figure}
\centerline{\hbox{\vbox{\psfig{figure=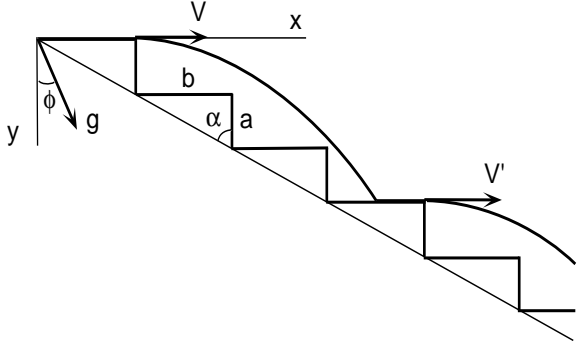,width=3.0true in,height=1.8true in}}}}
\vspace{0.5true cm}
\caption{Model for a single particle moving under gravity on an rough 
inclined surface.}
\label{fig:1}
\end{figure}

In our model, which is shown in Fig.\ \ref{fig:1}, the rough surface
is considered to have a simple staircase shape whose steps have height
$a$ and length $b$. For convenience, we choose a system of coordinates
in such a way that the step plateaus are aligned with the $x$ axis and
the direction of the acceleration of gravity {\bf g} makes an angle
$\phi$ with the $y$ axis.  A grain is then imagined to be launched on
the top of the `staircase' with a given initial velocity. In what
follows, we will be concerned with the problem of a {\it point}
particle falling down this `staircase' and will thus not take into
account any effect due to the finite size of the grain.  Upon reaching
the end of a step plateau, the particle will undergo a ballistic
flight until it collides with another plateau located a certain number
$n$ of steps below the departure step (e.g., $n=3$ in Fig.\ 1).
Accordingly, we will refer to the integer $n$ as the {\it jump number}
associated with this flight.

We will assume, for simplicity, that the momentum loss due to
collisions is determined by two coefficients of restitution $e_t$ and
$e_n$, corresponding to the tangential and normal directions,
respectively. More precisely, if ${\bf v} =(v_x,v_y)$ denote,
respectively, the components of the particle velocity parallel and
perpendicular to the surface  before a collision, then we will
take the velocity ${\bf v'} = (v_x',v_y')$ after the collision to be
given by
\begin{mathletters}
\label{eq:C}
\begin{eqnarray}
v'_x &=& e_t v_x , \label{eq:en}\\
v'_y &=&  - e_n v_y, \label{eq:et}
\end{eqnarray}
\end{mathletters}
where  $0\le e_t <1$ and $0\le e_n<1$. 

In the present paper we will for simplicity discuss only the case
$e_n=0$; the advantage being that the model can then be described by a
one-dimensional map. When $e_n > 0$ the dynamics is governed by a
three-dimensional map, the analysis of which is more complicated and
will be left for forthcoming publications \cite{us}.

We now derive the equations governing the dynamics of the model
presented above. Let us denote by $E$ the kinetic energy of the
particle at the moment of departure for a given flight. We write
$E={1\over2}mV^2$, where $m$ is the particle mass and $V$ is the
launching velocity at the start of the flight (see Fig.\
\ref{fig:1}). After this flight the particle will first collide with
a step below, then slide along this step (recall $e_n=0$), and
finally take off again on another flight with initial kinetic energy
$E'$. We suppose that the main energy loss is due to collisions
and so we neglect the energy dissipation as the particle slides along
a step, where it then moves with a constant acceleration $g\sin \phi$.
Using simple arguments of energy conservation together
with the collision conditions (1) and (2), one can write $E'$ in terms of $E$. 
The result is 
\begin{equation}
E'= {1\over2}m  e_t^2 v_x^2 + mg\sin\phi \, (nb-x) ,
\label{eq:2}
\end{equation}
where $n$ is the corresponding jump number for the flight and $x$ is
the $x$-coordinate of the landing point.  It takes a simple algebra to
show that at the landing point $(x,y)$ we have the following
identities:
\begin{mathletters}
\label{eq:3}
\begin{eqnarray}
&&x = {g \sin\phi\over 2} T^2 + \sqrt{{2E\over m}} T \label{eq:3a}, \\
&&y = {g \cos\phi\over 2} T^2 = na, \label{eq:3b} \\
&&v_x= g \sin\phi \, T + \sqrt{{2E\over m}}, \\
&&v_y= g \cos\phi \, T, 
\end{eqnarray}
\end{mathletters}
where $T$ is the flight time.

It is convenient to introduce a dimensionless energy-like variable:
\begin{equation}
{\cal E}={E\over{m g a \cos\phi}} . 
\label{eq:5}
\end{equation}
Eliminating $T$ from (\ref{eq:3}) and inserting the result into (\ref{eq:2}), 
we obtain that the dynamics of the model in terms of the variable 
${\cal E}$ is given by the following map:
\begin{equation}
{\cal E}' = f({\cal E},n)= n \left[ e_t^2 \, \left(\sqrt{{\cal E}/n}+ t\right)^2 + t 
\left(\tau -t - 2\sqrt{{\cal E}/n}\right)\right] .
\label{eq:6}
\end{equation}
where we have for conciseness introduced the notation
\begin{eqnarray}
&&t=\tan \phi, \\
&& \tau=b/a.
\end{eqnarray}
The parameter $\tau$ above can be viewed as a measure of the surface
roughness, with $\tau^{-1}=0$ corresponding to a perfectly smooth
surface. As for the inclination parameter $t$, we need to consider
only the interval $0<t<\tau$ for which non-trivial motion
occurs. (Clearly, for $t<0$ the particle will always come to a rest,
whereas for $t>\tau$ the particle undergoes a free fall without ever
colliding again with the ramp.)

\begin{figure}
\centerline{\hbox{\vbox{\psfig{figure=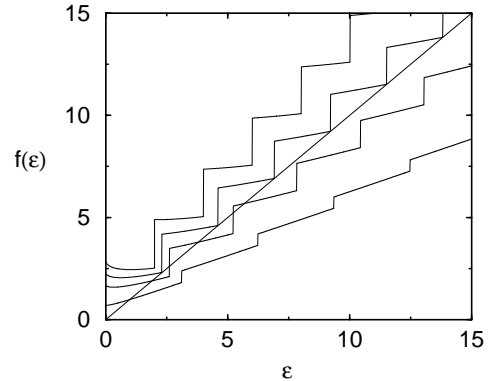,width=3.0true in,height=2.5true in}}}}
\vspace{-0.5true cm}
\caption{One-dimensional map $f({\cal E},n)$ for $e_t=0.7$, $\tau=3.7$
 and $t=$ 0.2, 0.5, 0.7, 0.9 (from the bottom up).} 
\label{fig:map}
\end{figure}

The flight jump number $n$ appearing in Eq.\ (\ref{eq:6}) is
determined from the energy $\cal E$ according to the following
condition: $n$ is equal to the smallest integer such that $nb-x\geq 0$
or, alternatively,
\begin{equation}
n(\tau-t) - 2\sqrt{n {\cal E}}\geq 0  .
\label{E-is-pos}
\end{equation} 
This means that ${\cal E}$ falls within the interval $I_n$:
\begin{equation} 
{\cal E} \in I_{n}(t) \equiv \left( {1\over4}(n-1) (\tau -  t)^2, 
{1\over4} n (\tau -  t)^2\right].
\label{eq:8}
\end{equation}
Thus the function $f({\cal E},n)$, as defined by Eqs.\ (\ref{eq:6})
and (\ref{eq:8}), exhibits jump discontinuities at energy values
${\cal E}= {1\over4} n \left(\tau-t\right)^2$, but each of its
branches is smooth.  This is illustrated in Fig.\
\ref{fig:map}, where we graph the function (\ref{eq:6}) for
$e_t=0.7$, $\tau=3.7$, and several values of the inclination $t$,

For later use, we note here that the average velocity $\overline{V}$
between two consecutive flights is given by
\begin{equation}
\overline{V}= {n L\over{T+ (\sqrt{2E'/m}-e_t v_x)/g\sin\phi}}, 
\label{eq:Vmean}
\end{equation}
where $L=\sqrt{a^2+b^2}$ and the second term in the denominator
corresponds to the time during which the particle moves on the ramp
(see Fig.\ 1).  If we now introduce a dimensionless mean velocity
\begin{equation}
\overline{{\cal V}}={\overline{V}\over\sqrt{a g
\cos\phi}}, \label{eq:Vdimless}
\end{equation}
then Eq.\ (\ref{eq:Vmean}) becomes
\begin{equation}
\overline{\cal V}= {t \sqrt{n (1+\tau^2)/2 }\over{(1-e_t)t + 
\sqrt{{\cal E}'/n}-e_t \sqrt{{\cal E}/n}}}. \label{eq:vmean}
\end{equation}

In order to study the dynamical properties of the map above, we must first
investigate the existence of fixed points.  If we denote by ${\cal E}_n$ a
fixed point with a jump number $n$, then ${\cal E}_n$ will be the
solution to the equation
\begin{equation}
{\cal E}_n = f({\cal E}_n,n) .\label{eq:fp}
\label{hmap:ex}
\end{equation}
In view of the homogeneity of the function $f({\cal E},n)$ [see Eq.\
(\ref{eq:6})] we write
\begin{equation}
{\cal E}_n = n [z_0(t)]^2, \label{eq:En}
\end{equation}
where the quantity $z_0(t)$ no longer bears any dependence on $n$.
Using Eqs.\ (\ref{eq:6}) and (\ref{eq:En}), Eq.\ (\ref{eq:fp}) becomes
\begin{equation}
(z_0+t)^2 = e_t^2 (t+z_0)^2 + \tau t, 
\label{eq114}
\end{equation}
whose positive solution is
\begin{equation}
z_0(t) = -t+\sqrt{{\tau t \over 1-e_t^2}}. \label{eq:z0}
\label{eq115}
\end{equation}

Now a fixed point ${\cal E}_n$, as given in Eqs.\ (\ref{eq:En}) and
(\ref{eq115}), will exist if and only if ${\cal E}_n\in I_n(t)$, where
the interval $I_n(t)$ is defined in (\ref{eq:8}).  Thus, as $t$
increases, a fixed point with jump number $n$ will be born when ${\cal
E}_n$ equals the left endpoint of $I_n$. Comparing Eqs.\ (\ref{eq:8}),
(\ref{eq:En}) and (\ref{eq:z0}), we see that this happens at an
inclination $t_n$ such that
\begin{equation} 
z_0(t_n)=-t_n+\sqrt{{\tau t_n \over 1-e_t^2}} = {1\over2} \sqrt{1-{1\over n}}
\left({\tau - t_n}\right) .
\label{eq116} 
\end{equation} 
This equation is quadratic in $\sqrt{t_n}$ and can thus be easily
solved. However, we shall not bother to give the result here and will
simply mention a few important facts that follow from Eq.\
(\ref{eq116}). First, we note that $t_1=0$ so that a fixed point with
jump number $n=1$ is always born at $t=0$.  Then, as $t$ increases,
fixed points with successively higher $n$ will appear in an increasing
sequence of inclinations $\{t_n\}_{n=1}^{\infty}$. Finally, we have
that for $t>t_\infty$, where $t_\infty = \lim_{n\to\infty} t_n$, all
fixed points cease to exist. Setting $n=\infty$ in Eq.\ (\ref{eq116})
we obtain for the limit point $t_\infty$:
\begin{equation} 
t_\infty = \tau {{1-e_t}\over{1+e_t}}.
\label{eq117} 
\end{equation}
The appearance of this sequence of fixed points can perhaps be best
visualized by referring to Fig.\ \ref{fig:map}, where we plot the
function $f({\cal E},n)$ at increasing values of $t$, with $e_t$ and
$\tau$ kept fixed.  For small $t$ (lower-most curve in Fig.\
\ref{fig:map}) there is only one intersection with the $45^\circ$
line, corresponding to the fixed point with $n=1$.  As $t$ increases
fixed points with successively higher $n$ appear (second curve from
the bottom).  At $t=t_\infty$ there are infinitely many such fixed points
(second curve from the top) and after this  all of them cease to
exist (uppermost curve).

One can also show that for $t>t_\infty$ we always have $f({\cal
E},n)>{\cal E}$, whereas for $0<t<t_\infty$ there exists an energy
${\cal E}^*$ such that $f({\cal E},n)<{\cal E}$ for ${\cal E}>{\cal
E}^*$ (see, e.g., Fig.\ \ref{fig:map}). We thus conclude that for
$t>t_\infty$ the particle velocity will become unbounded for any
initial condition, whereas for $0<t<t_\infty$ the velocity remains
always bounded. In other words, at the critical inclination
$t=t_\infty$ there is a sharp transition (independent of initial
conditions) from a regime of bounded velocity to accelerated motion.
In the region of bounded velocity, several dynamical regimes are
possible, depending on the stability of the fixed points, as discussed
below.

The stability of a fixed point ${\cal E}_n$ is determined by the
parameter $\lambda= f'({\cal E}_n,n)$, where the prime denotes
derivative with respect to $\cal E$, so that if $|\lambda|<1$
($|\lambda|>1$) the fixed point is stable (unstable) \cite{ott}.
Using Eqs.\ (\ref{eq:6}), (\ref{eq:En}) and (\ref{eq:z0}), we obtain
for the derivative $\lambda$ at the fixed point:
\begin{equation} 
\lambda(t) = 1 - {{1-e_t^2} \over 1 - 
\sqrt{{(1-e_t^2)t/\tau }}} .
\label{eq118} 
\end{equation}
Notice that $\lambda$ does not depend on $n$, thus implying that all
existing fixed points ${\cal E}_n$ (for given values of the model
parameters) have the same stability properties.  Moreover, since
$\lambda$ is always smaller than unity, we see that instability can occur
only if $\lambda (t)<-1$. Let us then denote by $t_{\rm inst}$ the
inclination such that $\lambda(t_{\rm inst})=-1$. From
Eq. (\ref{eq118}) we obtain that
\begin{equation}
t_{\rm inst}=\tau {(1+e_t^2)^2 \over 4(1-e_t^2)} .
\label{eq119}
\end{equation}
Thus  the fixed points are  stable for $t<t_{\rm inst}$ and unstable
for $t>t_{\rm inst}$.

If the fixed points are stable, the dynamics of the map will in
general be attracted to one of the existing fixed points. For example,
in the region of parameters such that $0<t<t_2<t_{\rm inst}$ the
particle will almost always reach a periodic motion where the particle
falls by one step at a time, since in this case only the fixed point
with $n=1$ exists and is stable \cite{note}. On the other hand, for
$t_2<t<t_{\rm inst}$ there are co-existing stable fixed points, in
which case the final state (i.e., the fixed points to which the
dynamics is attracted) will depend on the initial condition. Once the
system has reached a given fixed point ${\cal E}_n$ the particle will
accordingly be moving with a constant mean velocity $\overline{\cal
V}_n$ whose value can be readily obtained by inserting Eqs.\
(\ref{eq:En}) and (\ref{eq:z0}) into Eq.\ (\ref{eq:vmean}):
\begin{equation}
\overline{{\cal V}}_n=\left[{n(1+\tau^2)t\over{2 t_\infty}}\right]^{1/2} .  
\label{eq:Vn}
\end{equation}
 
\begin{figure}
\centerline{\hbox{\vbox{\psfig{figure=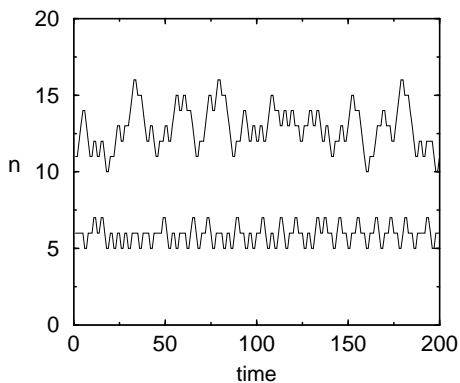,width=3.0truein,height=2.5true in}}}}
\vspace{-0.5true cm}
\caption{The jump number $n$ as a function of time (measured in iteration 
steps) in the chaotic regime. Here $e_t=0.35$, $\tau=3.73$, and
$t=1.65$ (lower orbit), 1.74 (upper orbit).}
\label{fig:chaos}
\end{figure}

When the fixed points are unstable ($t_{\rm inst}<t<t_\infty$), the
particle motion becomes very irregular and no stationary (periodic)
regime is ever reached. This is illustrated in Fig.\ \ref{fig:chaos},
where we plot the jump number $n$ as a function of time (iteration
step) for two orbits in the region where the fixed points are
unstable.  In this figure we clearly see that the jump number
fluctuates erratically around a mean value.  We have computed the
Lyapunov exponent for several values of parameters in the region of
unstable fixed points and have found it to be positive for all cases
studied, thus indicating that the motion is indeed chaotic in this
region.
 
\begin{figure}
\vspace{-2.0true cm}
\centerline{\hbox{\vbox{\psfig{figure=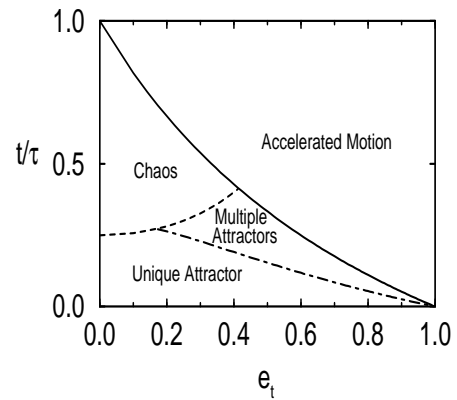,width=3.0true in,height=3.5true in}}}}
\vspace{-0.7true cm}
\caption{Phase diagram for the model.  The solid line corresponds to 
$t_\infty$, the dashed line to $t_{\rm inst}$, and the dot-dashed line
to $t_2$.}
\label{fig:phase}
\end{figure}

The different dynamical regimes displayed by the model above can be
conveniently summarized in terms of a ``phase diagram'' in the
parameter space $(e_t,t/\tau)$, as shown in Fig.\ \ref{fig:phase}. In
this figure we plot the curves corresponding to $t_\infty$ (solid
line) and $t_{\rm inst}$ (dashed line) given by Eqs.\ (\ref{eq117})
and (\ref{eq119}), respectively. Also plotted is the curve
representing the inclination $t_2$ (dot-dashed line) at which the
fixed point with $n=2$ first appears. Thus in terms of the
existence/stability of the fixed points the model displays the
following four regions: (i) for $0<t<{\rm min}(t_2,t_{\rm inst})$
there is a unique stable fixed point, namely, that with $n=1$; (ii)
for $t_2<t<{\rm min}(t_{\rm inst},t_\infty)$ there are multiple stable
fixed points (at least those with $n=1$ and $n=2$); (iii) for $t_{\rm
inst}<t<t_\infty$ all existing fixed point are unstable and chaotic
motion is observed; (iv) for $t>t_\infty$ no fixed point exists and
the motion becomes accelerated.  

Another interesting feature in Fig.\ \ref{fig:phase} is the fact that
the chaotic regime appears when the collisions are highly inelastic 
(i.e., small $e_t$). In particular, for $e_t>\sqrt{2}-1$ (at which point
$t_{\rm inst}$ equals $t_\infty$) the fixed points remain stable over
their entire domain of existence. (The results shown in Fig.\
\ref{fig:phase} are qualitatively different from the behavior seen in
the model studied in Ref.\ \cite{valance}, where chaotic motion
appears as the restitution coefficient increases.)

\section{Comparison with experiments}

\begin{figure}
\centerline{\hbox{\vbox{\psfig{figure=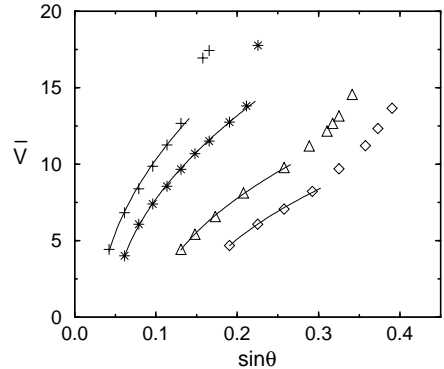,width=3.0true in,height=2.5true in}}}}
\vspace{-0.5true cm}
\caption{Mean velocity $\overline{V}$ (cm/s) as a function of 
$\sin \theta$. Points are experimental data taken from Ref.\ [4] for
$R/r=2$ ($+$), 1.5 ($\ast$), 1 ($\triangle$), 0.8 ($\Diamond$). Solid
curves are theoretical fits [Eq.\ (\ref{eq:V1})], ending near
the last data point considered in the fit. Fitted parameters are
$(e_t,\tau)$ = (0.72, 33.18), (0.64, 21.09), (0.41, 10.17),
(0.27, 7.07), from left to right.}
\label{fig:fit}
\end{figure}

In this section we wish to compare our model with recent experimental
studies of a single ball moving under gravity on a rough inclined
surface. In these experiments, first performed by Jan {\it et al.}
\cite{jan} and later expanded by Ristow {\it et al.}  \cite{ristow}, a
rough surface was constructed by gluing steel spheres of radius $r$ on
a L-shaped flume. Another steel sphere of radius $R$ was then launched
with a small initial velocity and its subsequent motion analyzed. As
the surface inclinations increases, the following three regimes are
observed \cite{ristow}: for small inclinations the bead always stops
(regime A), then comes a range of inclinations for which the ball
reaches a steady state with constant mean velocity (regime B), and
beyond this point the ball starts to jump (regime C).  In Fig.\
\ref{fig:fit} we show data taken from Ref.\ \cite{ristow} for the ball
mean velocity $\overline{V}$ as a function of $\sin \theta$, where
$\theta$ is the inclination angle with respect to the horizontal
direction. As discussed in Ref.\ \cite{ristow}, the change in trend
observed in the data as $\theta$ increases (for a given value of
$R/r$) marks the beginning of the jumping regime.

The regime B seen in the experiments corresponds in our model to a
stable fixed point with $n=1$, for in this case the particle reaches a
periodic motion where it falls one step at a time (as in the
experiments).  In order to compare our model more closely with the
experiments let us first express the mean velocity $\overline{V}_1$
(at the fixed point $n=1$) in terms of the angle $\theta$, where
$\theta=\phi+\pi/2-\alpha$ (see Fig.\ 1).  Setting $n=1$ in Eq.\
(\ref{eq:Vn}), returning to dimensionful units via Eq.\
(\ref{eq:Vdimless}), and expressing the final result in terms of
$\theta$, we  obtain
\begin{equation}
\overline{V}_1=\left[{L g (1+e_t)\over{2(1-e_t)}}\right]^{1/2}
\sqrt{\sin\theta-\tau^{-1} \cos\theta}. \label{eq:V1}
\end{equation}
(We remark parenthetically that a similar expression can be obtained
heuristically if one introduces an effective sliding friction in
addition to inelastic collisions; see Refs. \cite{jan,ristow}. Our
formula follows however from a pure collision model.)

We have fitted the expression (\ref{eq:V1}) to the experimental data
shown in Fig.\ \ref{fig:fit} --- the corresponding results being
displayed as solid curves in this figure. In our fitting procedure, we
took $L=2r=1$ cm \cite{ristow}, $g=980$ cm/s$^2$, and best-fitted the
parameters $\tau$ and $e_t$ for each data set considering {\it only}
points in regime B. As we see in Fig.\ \ref{fig:fit}, the model
prediction for the dependence of $\overline{V}$ with $\theta$ is in a
good agreement with the experimental data (in regime B).

The jumping regime observed in the experiments, on the other hand,
would correspond in our model to the region of unstable fixed points,
since in this case the particle jumps erratically never reaching a
steady state (see Fig.\ \ref{fig:chaos}). This analogy might then
provide a possible explanation for the change in trend observed in the
experimental data for large inclinations.  To see
this, consider the region of small $e_t$ in the phase diagram shown in
Fig.\ \ref{fig:phase}. As the inclination $t$ increases (for a given
$e_t$) the system goes from a region of stable periodic motion (with $n=1$)
to a regime of chaotic jumps, in close resemblance to the experimental
transition from steady-state to the jumping regime.

\begin{figure}
\centerline{\hbox{\vbox{\psfig{figure=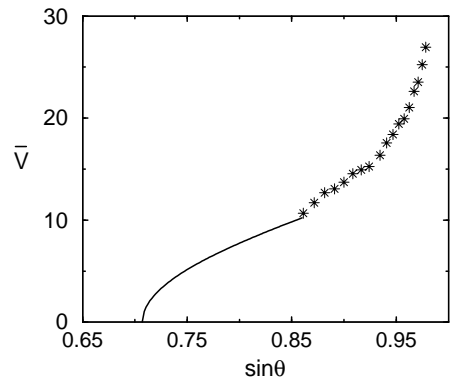,width=3.0true in,height=2.5true in}}}}
\vspace{-0.5true cm}
\caption{Same as figure \ref{fig:fit} for our model with $e_t=0.1$ and 
$\tau=1$. The solid curve corresponds to Eq.\ (\ref{eq:V1}) whereas the
stars give the computed mean velocity in the chaotic regime.}
\label{fig:vmean}
\end{figure}

To probe this analogy further, we illustrate in Fig.\ \ref{fig:vmean}
the behavior predicted by the model for the mean velocity
$\overline{V}$ as a function of $\sin \theta$ in the region of small
$e_t$. In this figure, the solid curve corresponds to the expression
(\ref{eq:V1}) for $\overline{V}_1$, up to the point where the fixed
point goes unstable, and the crosses are computed values of
$\overline{V}$ in the ensuing chaotic regime. Comparing Fig.\
\ref{fig:vmean} with Fig.\ \ref{fig:fit}, we see that the change in
behavior predicted by the model at the onset of instability is in
qualitative agreement with what is observed in the experiments (for
small values of $R/r$) as the ball enters the jumping regime.  Of
course, more detailed experiments are necessary to verify whether
chaotic motion does indeed take place in the jumping regime.

\section{Conclusions}

We have studied a simple geometrical model for the gravity-driven
motion of a single particle on a rough inclined line. In our model the
rough line was chosen to have a regular staircase shape and a simple
collision law was adopted. With these simplifications the dynamics is
described by a one-dimensional map that is quite amenable to
analytical treatment.  Summarizing our findings, we have seen that our
model displays the following four dynamical regimes:
\begin{enumerate}
\item for $0<t<{\rm min}(t_2,t_{\rm inst})$ there is a unique stable
fixed point.
\item for $t_2<t<{\rm min}(t_{\rm inst},t_\infty)$ the system has
multiple stable fixed points.
\item for $t_{\rm inst}<t<t_\infty$ the fixed points are unstable
and the dynamics is  chaotic.
\item for $t>t_\infty$ no fixed point exists and the motion becomes 
accelerated.
\end{enumerate}
Here the parameter $t$ measures the surface inclination and the
quantities $t_2$, $t_{\rm inst}$, and $t_\infty$ separating the
different regimes are given in terms of the other two model-parameters, namely,
the restitution coefficient $e_t$ and the roughness parameter
$\tau$. These regimes are indicated in the phase diagram shown in
Fig.\ \ref{fig:phase}.  Furthermore, it can be shown
\cite{companion} that the above conclusions, which were derived in the
context of a simple collision rule, remain valid for a wide class of
tangential restitution laws.

Despite its simplicity, our model does provide a theoretical framework
within which the generic behavior seen in experiments on a ball
moving on a rough surface can be qualitatively understood. For
example, the model successfully predicts the existence of several
dynamical regimes that are also observed in the experiments.  In
particular, the predicted functional dependence of the mean velocity
with the inclination angle $\theta$ (in the steady-state regime) is in
good agreement with the experiments. Moreover, the model provides a
possible explanation for the change in trend seen in the experimental
data as the ball enters the jumping regime. We have suggested that
this jumping regime might correspond to a chaotic motion, as so
happens in the model. Clearly, more experimental studies are required
to investigate this interesting possibility.

This work was supported in part by FINEP and CNPq.


\begin{references}
\bibitem{note} When $t$ is very close to $t_{\rm inst}$ the system may
enter a chaotic regime for certain initial conditions, while
converging to the fixed point for others.
\end{references}
\end{document}